\documentclass[%
 reprint,
 amsmath,amssymb,
 aps,
pra,
]{revtex4-2}

\usepackage{graphicx}
\usepackage{dcolumn}
\usepackage{bm}
\usepackage{hyperref}

\begin{document}

\title{Generalized Einstein Relations between Absorption and Emission Spectra at Thermodynamic Equilibrium}

\author{Jisu Ryu}
\thanks{These authors contributed equally to this work}
\author{Sarang Yeola}
\thanks{These authors contributed equally to this work}
\author{David M. Jonas}
\email[Correspondence to: ]{jonasd@colorado.edu}
\affiliation{University of Colorado, Boulder, CO 80309, USA}

\begin{abstract}
We present Einstein coefficient spectra and a detailed-balance derivation of generalized Einstein relations between them that is based on the connection between spontaneous and stimulated emission. If two broadened levels or bands overlap in energy, transitions between them need not be purely absorptive or emissive. Consequently, spontaneous emission can occur in both transition directions, and four Einstein coefficient spectra replace the three Einstein coefficients for a line. At equilibrium, the four different spectra obey five pairwise relationships and one lineshape generates all four. These relationships are independent of molecular quantum statistics and predict the Stokes’ shift between forward and reverse transitions required by equilibrium with blackbody radiation. For Boltzmann statistics, the relative strengths of forward and reverse transitions depend on the formal chemical potential difference between the initial and final bands, which becomes the standard chemical potential difference for ideal solutes. The formal chemical potential of a band replaces both the energy and degeneracy of a quantum level. Like the energies of quantum levels, the formal chemical potentials of bands obey the Rydberg-Ritz combination principle. Each stimulated Einstein coefficient spectrum gives a frequency-dependent transition cross section. Transition cross sections obey causality and a detailed-balance condition with spontaneous emission, but do not directly obey generalized Einstein relations. Even with an energetic width much less than the photon energy, an absorptive forward transition with an energetic width much greater than the thermal energy can have such an extreme Stokes’ shift that its reverse transition cross section becomes predominantly absorptive rather than emissive.
\end{abstract}


\date{\today}

\maketitle

Einstein's relationships between single-photon absorption, stimulated emission, and spontaneous emission \cite{RN978,RN979,RN1457,RN1458} conflict with the time-energy uncertainty principle\cite{RN1417} by ascribing a finite lifetime to the upper state of an infinitely narrow spectroscopic line. Since all quantum levels are radiatively broadened,\cite{RN3900} a generalization of Einstein's treatment is needed. This paper presents an internally consistent treatment of the thermal equilibrium relationships between absorption, stimulated emission, and spontaneous emission that obeys detailed balance and causality and is compatible with the time-energy uncertainty principle.

\begin{figure}[hp]
\centering
\includegraphics[width=\linewidth]{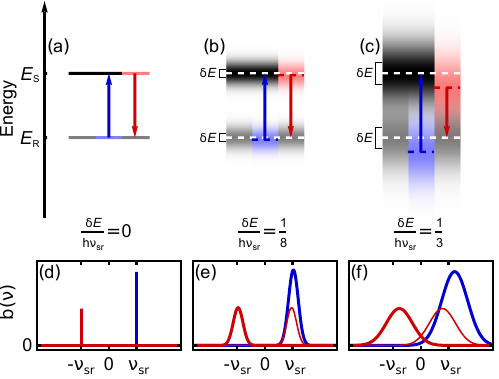}
\caption{Transitions between two bands for three different amounts of band broadening. In all three columns, the thermal energy is $k_{\text{B}}T$/$h\nu_{sr} =$ 1/2 and the density of states ratio is $\rho_{S}$:$\rho_{R} =$ 2:1. In each column, both bands have the same Gaussian band broadening with standard deviation $\delta E$. (a) Infinitely narrow bands or quantum energy levels as in Einstein's theory for line spectra; $\delta E$ /$h\nu_{sr} =$ 0. (b) Broadened bands in which energetic overlap between bands can be practically neglected, as in McCumber's relationship between cross sections; $\delta E$ /$h\nu_{sr} =$ 1/8. (c) The general case of energetically overlapping bands with widths comparable to the thermal energy treated here; $\delta E$ /$h\nu_{sr} =$ 1/3. Within each column, the density of states for both bands is shown in grayscale at left, the $R$ to $S$ transition is in the middle (upward blue arrow from blue thermal population distribution in $R$ to gray density of states in $S)$, and the $S$ to $R$ transition is at right (downward red arrow from red thermal population distribution in $S$ to gray density of states in $R)$. $E$ is the molecular energy. In each column, the lower panels (d), (e), and (f) show the corresponding spectra of signed Bohr transition frequencies arising from double-convolution of the conditional thermal population distribution in the initial band and the density of states in the final band with a photon $+$ molecule energy-conserving delta function. Thick blue curves show the spectrum for the transition from $R$ to $S$ (mostly absorption at positive transition frequencies) and thick red curves show the spectrum for the transition from $S$ to $R$ (mostly emission at negative transition frequencies). To highlight the Stokes' shift, thin red curves show the $S$ to $R$ spectrum on a frequency-reversed axis. In (a) and (d), the absorption and emission photon energies are equal. In (b) and (e), the frequency-reversed emission spectrum is centered slightly below the overlapping absorption spectrum (Stokes' shift). In (e), $R$ to $S$ transition frequencies are practically confined to positive frequencies (absorption) and $S$ to $R$ transition frequencies are practically confined to negative frequencies (emission). In (c) and (f), there is a larger Stokes' shift and some configurations within band $S$ lie below thermally populated configurations within band $R$, so that the spectrum of $R$ to $S$ transition frequencies in (f) extends across zero frequency, involves both absorption and emission, and is unclassifiable as either. The spectrum of $S$ to $R$ transition frequencies in (f) extends even further across zero frequency, and is also unclassifiable as either emission or absorption.}
\label{fig:1}
\end{figure}

Beyond practical use of the same lineshape for narrow absorption and emission transitions,\cite{RN1457,RN1458,RN3162} all detailed-balance attempts to generalize Einstein's absorption-emission relations treat only one of two essential difficulties. First, transitions with widths comparable to the average transition photon energy create the difficulty that the range of final--initial energy differences spreads across zero; Van Vleck, Weisskopf, and Margenau treated such transitions in the limit of width very much less than the thermal energy.\cite{RN3500,RN3499} Second, transitions with widths comparable to the thermal energy create the difficulty that equilibrium within the initial level affects absorption and emission differently; McCumber treated such transitions,\cite{RN2267} but his treatment has previously unstated restrictions that limit the width compared to the transition photon energy and the thermal energy. Our prior introduction of 3 Einstein coefficient spectra for transitions between two broadened levels\cite{RN2971} has the same unstated restrictions as McCumber's. The combination of both difficulties is illustrated in Fig. \ref{fig:1}, which also shows spectra of possible transition frequencies for increasing level widths. Prior attempts to generalize have not simultaneously treated both essential difficulties, nor have they demonstrated, as Einstein\cite{RN978} and Milne\cite{RN3907} did, that equilibrium with blackbody radiation drives molecular translational equilibrium.

The derivation of generalized Einstein relations between Einstein coefficient spectra presented here uses detailed balance and quantum properties of light to obtain powerful thermodynamic relationships between spectra - molecular quantum and statistical mechanics are not used. In particular, molecular energy levels and Bohr transition frequencies play no role in the derivation and transition frequencies are not assumed to obey the Bohr frequency condition. For infinitely narrow levels, a molecular Boltzmann distribution, and Einstein's quantum conditions, the generalized Einstein relations give Einstein's results and the Bohr frequency condition. In this paper, we use the phrase Bohr transition frequency and simplified quantum models only to motivate the form of the spectra, illustrate how quantum results can be used in a kinetic and thermodynamic theory of spectroscopy, and argue that an extreme consequence of the generalized Einstein relations is necessary. \vspace{54pt}

\section*{Hypotheses}
Here, we treat thermal equilibrium transitions between bands in molecules. Molecules may be any finite-sized single-photon absorber made up of bound particles: an atom in vacuum, a molecule in solution, a protein (even one containing multiple pigments), a single many-body system, \textit{etc.} Einstein's infinitely narrow quantum levels are generalized to broadened molecular bands. Each band has a thermodynamic equilibrium population and must encompass coherent molecule-environment evolution during radiative transitions so that any single-photon transition ends within one band. Molecules often equilibrate among constituent forms that can be separately quantified but not physically separated.\cite{RN1791} We will treat a band as a thermodynamic constituent on a spectroscopic measurement timescale; the equilibrium properties of a band depend on thermodynamic properties such as temperature and pressure.

From a quantum perspective, each band incorporates coupled states that share a common characteristic or characteristics; from a thermodynamic perspective, those same characteristics partition the molecular population among bands.\cite{RN3938} Within each band, all remaining degrees of freedom for molecule, environment, and radiation are non-characteristic and freely variable. For example, with a molecular electronic state (including spin) as the common characteristic of a band, the accompanying vibrational, rotational, solvent, and radiation field degrees of freedom are non-characteristic and freely variable. If a band incorporates several electronic states (for example, by specifying the number of electrons $n$ and holes $p$ in a small piece of semiconductor), the electronic state within the band becomes non-characteristic and freely variable.

Unlike quantum levels, two broadened bands can overlap in energy so that a transition between them in one direction (for example, $R\to S$) can involve both absorption and emission; as a result, the molecular transition is unclassifiable as either. In such cases, we speak of forward and reverse molecular transitions. The energetic overlap between two bands that makes transitions between them unclassifiable in practice is common in transitions between excited electronic states of molecules and between excited bands in semiconductors. Since thermal excitations within a band (such as phonon or vibrational energy levels within an electronic band or state) often have no energetic upper bound, energetic overlap between bands is typical even when it is not practically important. Even if a molecular transition between two bands is unclassifiable, each single-photon transition between the two bands can still be classified as absorption or emission according to whether it annihilates or creates a photon. It is convenient to use the sign of the cyclic frequency to distinguish photon absorption ($\nu >0$) from photon emission ($\nu <0$). Since every broadened molecular transition can involve stimulated emission, every broadened molecular transition can also occur by spontaneous emission.

To accommodate energetically overlapping bands, we replace Einstein's set of three non-negative coefficients for an infinitely sharp spectroscopic line with a set of four non-negative Einstein coefficient spectra for transitions between 
two bands $R$ and $S$. The integrals of these spectra give Einstein coefficients:
\begin{subequations}
\label{eq1}
    \begin{align}
    \label{eq1a}
    \begin{split}
    & B_{S\to R} (p,T)=\int_{-\infty }^{+\infty } {b_{S\to R} (\nu,p,T)d\nu },
    \\ & \hspace{7em} \text{(stimulated transition from \textit{S} to \textit{R})}
    \end{split}\\
    \label{eq1b}
    \begin{split}
    & A_{S\to R} (p,T)=\int_{{\kern 1pt}0}^{+\infty } {a_{S\to R}^{\nu } (-\nu,p,T)d\nu },
    \\ & \hspace{7em}  \text{(spontaneous transition from \textit{S} to \textit{R})}
    \end{split} 
\end{align}
\begin{align}
    \label{eq1c} 
    \begin{split}
    & B_{R\to S} (p,T)=\int_{-\infty }^{+\infty } {b_{R\to S} (\nu ,p,T)d\nu },
    \\ & \hspace{7em}\text{(stimulated transition from \textit{R} to \textit{S})}
    \end{split} \\
    \label{eq1d}
    \begin{split}
    & A_{R\to S} (p,T)=\int_{{\kern 1pt}0}^{+\infty } {a_{R\to S}^{\nu } (-\nu,p,T)d\nu }.
    \\ & \hspace{7em}  \text{(spontaneous transition from \textit{R} to \textit{S})}
    \end{split}
    \end{align}
\end{subequations}

All four Einstein coefficient spectra depend on pressure $p$, temperature $T$, system composition, and external potentials or fields, but we have omitted system composition, external potentials, and fields from the notation for simplicity. We have avoided labeling the stimulated transitions as either absorption or emission. Spontaneous emission spectral densities have a right superscript $\nu $.\cite{RN3931} The transition in Eq. [\ref{eq1d}] would not occur for infinitely narrow levels with $S$ above $R$, but is appreciable for the situation in Fig. \ref{fig:1}c and \ref{fig:1}f. Finally, this approach includes intraband transitions within a single band, where $R = S$ and only two Einstein coefficient spectra exist.

The fundamental hypothesis of this paper assumes that the conditional transition probabilities per unit time for a molecule in band $S$ to make a single-photon transition to band $R$ are:
\begin{subequations}
\label{eq2}
\begin{flalign}
    \label{eq2a}
    \begin{split}
& { }^{b}\Gamma_{S\to R} (u_{+}^{\nu } ;p,T)  \\
& \quad =\int_{{\kern 1pt}0}^{+\infty } 
{[b_{S\to R} (\nu ,p,T)+b_{S\to R} (-\nu ,p,T)]
u_{+}^{\nu } (\nu )d\nu } ,
    \\ & \hspace{17em} \text{   (stimulated)}
    \end{split} \\
    \label{eq2b}
    \begin{split}
& { }^{a}\Gamma_{S\to R} (p,T)=\int_{{\kern 1pt}0}^{+\infty } {a_{S\to R}^{\nu } (-\nu ,p,T)d\nu } ,
    \\ & \hspace{17em} \text{ (spontaneous)}
    \end{split}
\end{flalign}
\end{subequations}

\noindent where $u_{+}^{\nu } (\nu )$ is the positive-frequency spectral density of electromagnetic energy per unit volume. In Eq. [\ref{eq2a}], the first product inside the integral represents absorption from $S$ to $R$ and the second product represents stimulated emission from $S$ to $R$. The use of conditional transition probabilities per unit time assumes weak molecule-field coupling. The simple form of Eq. [\ref{eq2}] assumes that molecules are isotropic or pseudo-isotropic through time-averaging\cite{RN978} (so that $a$ and $b$ are independent of electromagnetic polarization vector \mbox{\boldmath$\varepsilon$} and wavevector \textbf{k}) and assumes a homogeneous and isotropic medium.\cite{RN978} The total conditional transition probability per unit time for a single-photon transition from $S$ to $R$ is
\begin{equation}
\label{eq3}
\Gamma_{S\to R} (u_{+}^{\nu } ;p,T)={ }^{b}\Gamma_{S\to R} (u_{+}^{\nu } 
;p,T)+{ }^{a}\Gamma_{S\to R} (p,T).
\end{equation}
The same expressions, with band subscripts interchanged, hold for molecular transitions from $R$ to $S$. These expressions reduce to Einstein's for infinitely narrow lines.

Einstein's derivation of relationships for line spectra in vacuum explicitly supposed that the $A$ and $B$ coefficients are constants.\cite{RN978,RN3951} Since equilibrium bands and spectra depend on temperature, the generalized Einstein relations must be derived differently. A fundamentally different derivation is necessary even for radiatively broadened transitions of a single molecule in infinite vacuum because 
emission that is stimulated by temperature-dependent blackbody radiation dominates over spontaneous emission for frequencies $\nu <\ln (2)k_{\text{B}} T/h$.\cite{RN432} As a result, radiative lifetimes and radiative linewidths are temperature dependent,\cite{RN3162,RN3414} so that Einstein coefficient spectra always depend on temperature. Only the pressure dependence in Eqs. [\ref{eq1}]-[\ref{eq3}] disappears in vacuum.

We now consider what can be deduced at a single temperature and pressure from radiative transitions between a pair of broadened bands at equilibrium. At equilibrium, there is no distinction between the forward and backward direction of time for molecular processes.\cite{RN2662,RN3917,RN3915} This time-reversal 
invariance is necessary for equilibrium and underlies detailed balance.\cite{RN3918,RN3492} Detailed balance between time-reversed processes at equilibrium demands not only that the integrated rates from Eqs. [\ref{eq2}] and [\ref{eq3}] for radiative transitions from $S$ to $R$ balance the integrated rates for radiative transitions from $R$ to $S$, but further demands that, over any frequency interval, the rate for equilibrium total emission (spontaneous plus stimulated) from $S$ to $R$ must exactly balance the rate for its time-reversed process, which is equilibrium absorption from $R$ to $S$ over the same frequency interval. Because it requires time-reversal invariance, this detailed balance can be violated, for example, in a fixed external magnetic field or if the entire system is rotating.\cite{RN3492,RN3937}

For a single molecule, each time-averaged equilibrium rate is equal to the product of the equilibrium probability for the prior condition of occupying the initial band [for example, ${ }^{eq}\mathcal{P}_{S} (p,T)$ for band $S$] with the equilibrium conditional transition probability per unit time. For any bands $R$ and $S$, detailed balance between time-reversed processes at equilibrium equates the single-molecule, time-averaged equilibrium rate for total emission from $S$ to $R$ to the single-molecule, time-averaged equilibrium rate 
for absorption from $R$ to $S$:

\begin{align}
\label{eq4}
\begin{split}
& { }^{eq}\mathcal{P}_{S} [b_{S\to R} (-\nu )u_{\text{BB+}}^{\nu } (\nu )+a_{S\to 
R}^{\nu } (-\nu )]d\nu \\ 
& \quad ={ }^{eq}\mathcal{P}_{R} b_{R\to S} (\nu )u_{\text{BB+}}^{\nu } 
(\nu )d\nu ,
\end{split} 
\end{align}
where $u_{\text{BB+}}^{\nu } (\nu ,p,T)$ is the positive-frequency spectral density of blackbody radiation per unit volume and the frequency interval $d\nu $ can be as small as we like. Every quantity in Eq. [\ref{eq4}] is a function of the thermodynamic variables ($p$, $T$, \textit{etc.}) and these must be the same throughout but have been suppressed to emphasize the frequency where it appears. Similarly, the time-averaged equilibrium rate for total emission from $R$ to $S$ must equal the time-averaged equilibrium rate for absorption from $S$ to $R$, but this result is obtained from Eq. [\ref{eq4}] by exchanging band labels. Equation [\ref{eq4}] also applies to intraband transitions with $R = S$. Equation [\ref{eq4}] preserves both equilibrium band populations and equilibrium photon numbers and can also be 
derived by balancing both. Einstein's detailed-balance treatment\cite{RN978} appeared to be objectionable because it combined two different rate laws on one side but not the other,\cite{RN3915} as on the left and right of Eq. [\ref{eq4}]; this apparent inconsistency was first resolved by Bothe\cite{RN432} and his resolution will be needed for the derivation in the next section.

Solving Eq. [\ref{eq4}] for the equilibrium blackbody radiation gives
\begin{equation}
\label{eq5}
u_{\text{BB+}}^{\nu } (\nu ,p,T)=\dfrac{\left[ {\dfrac{a_{S\to R}^{\nu } (-\nu 
,p,T)}{b_{S\to R} (-\nu ,p,T)}} \right]}{\left[ {\dfrac{{ }^{eq}\mathcal{P}_{R} 
(p,T)b_{R\to S} (\nu ,p,T)}{{ }^{eq}\mathcal{P}_{S} (p,T)b_{S\to R} (-\nu ,p,T)}} 
\right]-1}.
\end{equation}
Assuming dilute molecules and that any cavity surrounding the sample is large (so that the density of modes becomes continuous),\cite{RN978} the Planck blackbody radiation spectral density may be written in terms of the positive frequency spectral density of electromagnetic modes per unit volume $G^{\nu }_{+}(\nu ,p,T)$:
\begin{equation}
\label{eq6}
u_{\text{BB+}}^{\nu } (\nu ,p,T)=\frac{h\nu G^{\nu }_{+}(\nu ,p,T)}{\exp (h\nu /k_{\text{B}} T)-1},
\end{equation}
where $h$ is the Planck constant and $k_{\text{B}} $ is the Boltzmann constant. 
For example, a linear, homogeneous, and isotropic sample that is weakly dispersive and (approximately) transparent has \cite{RN3920,RN3495}
\begin{equation}
\label{eq7}
G^{\nu }_{+}(\nu ,p,T)=8\pi \nu^{2}[n(\nu )]^{2}[\partial(\nu n(\nu ))/{\kern 1pt}\partial \nu ]\theta (\nu )/c^{3},
\end{equation}
where $c$ is the speed of light in vacuum, the refractive index $n$ depends on frequency, pressure, temperature, composition, \textit{etc.} ($p$, $T$, \textit{etc.} have been suppressed), and $\theta (\nu )$ is a Heaviside unit step function that restricts $\nu \ge 0$. If we assume that the Planck law describes radiation at thermal equilibrium, comparing Eqs. [\ref{eq5}] and [\ref{eq6}] immediately suggests the generalized Einstein relations:
\begin{align*}
a_{S\to R}^{\nu } (-\nu ,p,T)&=h\nu G^{\nu }_{+}(\nu ,p,T)b_{S\to R} (-\nu ,p,T),\\
b_{S\to R} (-\nu ,p,T)&=\frac{{ }^{eq}\mathcal{P}_{R} (p,T)}{{ }^{eq}\mathcal{P}_{S} (p,T)}b_{R\to S} (\nu ,p,T)
\exp (-h\nu /k_{\text{B}} T).
\end{align*}
However, additional physical considerations are necessary to justify detailed balance between absorption and stimulated plus spontaneous emission in Eq. [\ref{eq4}] and to establish these relations as the unique solution to Eq. [\ref{eq4}].

\section*{Derivation}

\begin{figure*}[t!]
\centering
\includegraphics[width=17.6cm,height=7.7cm]{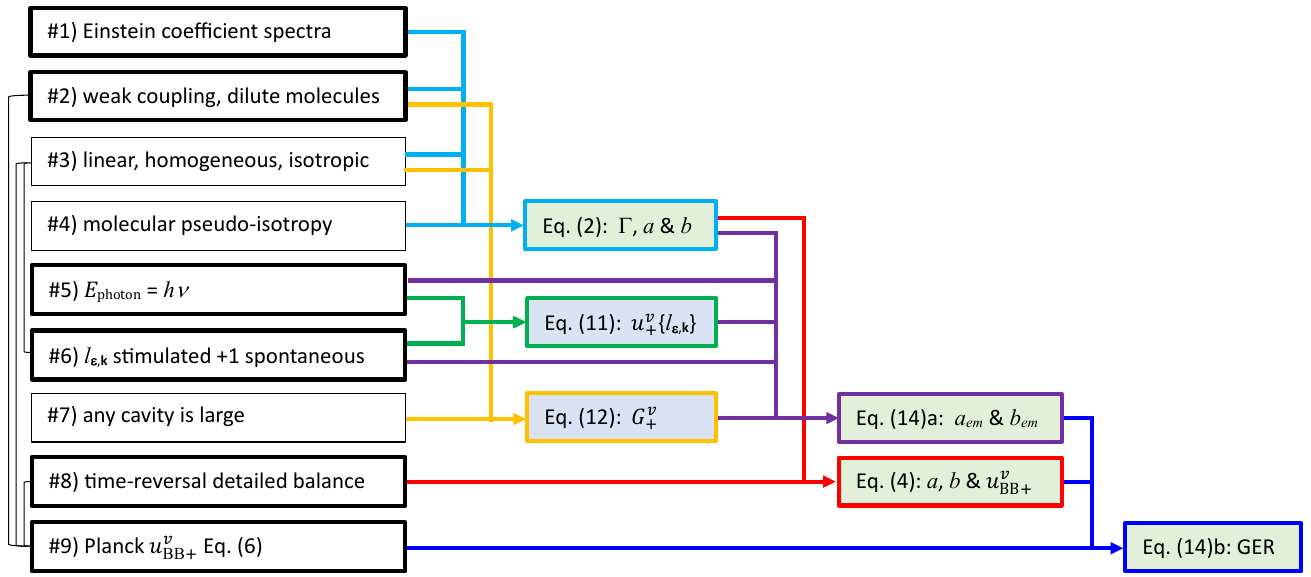}
\caption{Flow of detailed-balance derivation of the generalized Einstein relations. Nine assumptions are in black bordered boxes at left, with thick borders for the hypothesis and fundamental assumptions and thin borders for simplifying assumptions that could be modified. The tie bars at left indicate prior assumptions that are required for later assumptions. Left to right and top to bottom, the colored arrows show how assumptions combine to generate results in boxes with the same-colored border and how assumptions and prior results combine to generate further results. Light blue shading within a box indicates a previously known result.}
\label{fig:2}
\end{figure*}

Figure \ref{fig:2} shows the flow of a derivation that depends on two fundamental physical assumptions: First) that a photon has energy $E=h\nu $; Second) that stimulated and spontaneous emission are two aspects of a single emission process in which, for each mode
(\mbox{\boldmath$\varepsilon$},\textbf{k}) of the electromagnetic field, the conditional transition probability for emission is proportional to
$(l_{{\mbox{\footnotesize \boldmath$\varepsilon$}},{\rm {\bf k}}} +1)$, where $l_{{\mbox{\footnotesize \boldmath$\varepsilon$}},{\rm {\bf k}}} $ is the number of photons initially present in mode
$({\mbox{\boldmath$\varepsilon$}},{\rm {\bf k}})$.
Bothe\cite{RN432} identified the part of the conditional transition probability which is proportional to
$l_{{\mbox{\footnotesize \boldmath$\varepsilon$}},{\rm {\bf k}}} $ as stimulated emission and the part which is proportional to 1 as spontaneous emission. Spontaneous emission of a photon can occur into any electromagnetic mode with a transition probability that is independent of the number of photons initially present in that mode. Bothe's recognition that there is fundamentally one emission process [for example from
$(S,l_{{\mbox{\footnotesize \boldmath$\varepsilon$}},{\rm {\bf k}}} )$ to 
$(R,l_{{\mbox{\footnotesize \boldmath$\varepsilon$}},{\rm {\bf k}}} +1)$] with a transition probability proportional to
$(l_{{\mbox{\footnotesize \boldmath$\varepsilon$}},{\rm {\bf k}}} +1)$ 
for each mode\cite{RN432} was proven by Dirac\cite{RN903} and justified Einstein's apparently objectionable\cite{RN3915} step of equating the sum of the two emission rates to the absorption rate when invoking detailed balance. It justifies taking total emission from $S$ to $R$ as the time-reversal of absorption from $R$ to $S$ in Eq. [\ref{eq4}]. Fundamentally, this single-photon transition probability proportional to
$(l_{{\mbox{\footnotesize \boldmath$\varepsilon$}},{\rm {\bf k}}} +1)$ arises from a quantum electrodynamic treatment of the electromagnetic fields as linear harmonic oscillators.\cite{RN903,RN1419} This proportionality is common to all single-photon transitions, so there is no need to specify the molecule-field interaction further (in particular, the results do not depend on a multipole expansion, let alone a specific electric or magnetic multipole transition order).

Based on the second fundamental physical assumption above, the conditional transition probability per unit time for total emission involves sums of the form
\begin{align}
\begin{split}
\label{eq8}
 { }^{em}\Gamma_{S\to R}
 & =\sum\limits_{{\mbox{\footnotesize \boldmath$\varepsilon$}},{\rm {\bf 
k}},l} {\mathcal{P}_{{\mbox{\footnotesize \boldmath$\varepsilon$}},{\rm {\bf k}}} (l_{{\mbox{\footnotesize \boldmath$\varepsilon$}},{\rm {\bf k}}} )\beta_{S\to R} ({\mbox{\boldmath$\varepsilon$}},{\rm {\bf 
k}})(l_{{\mbox{\footnotesize \boldmath$\varepsilon$}},{\rm {\bf k}}} +1)} \\ 
 & =\sum\limits_{{\mbox{\footnotesize \boldmath$\varepsilon$}},{\rm {\bf k}},l} {\mathcal{P}_
{{\mbox{\footnotesize \boldmath$\varepsilon$}},{\rm {\bf k}}} (l_{{\mbox{\footnotesize \boldmath$\varepsilon$}},{\rm {\bf k}}} 
)\beta_{S\to R} ({\mbox{\boldmath$\varepsilon$}},{\rm {\bf k}})l_{{\mbox{\footnotesize \boldmath$\varepsilon$}},{\rm {\bf k}}} } +\sum\limits_{{\mbox{\footnotesize \boldmath$\varepsilon$}},{\rm 
{\bf k}}} {\beta_{S\to R} ({\mbox{\boldmath$\varepsilon$}},{\rm {\bf k}})} ,
\end{split}
\end{align}
where $\mathcal{P}_{{\mbox{\footnotesize \boldmath$\varepsilon$}},{\rm {\bf k}}} (l_{{\mbox{\footnotesize \boldmath$\varepsilon$}},{\rm {\bf k}}} )$ is the probability that mode
$({\mbox{\boldmath$\varepsilon$}},{\rm {\bf k}})$ with positive frequency $\nu_{{\rm {\bf k}}} $ contains 
$l_{{\mbox{\footnotesize \boldmath$\varepsilon$}},{\rm {\bf k}}} $ photons before the emission 
transition and $\beta_{S\to R} ({\mbox{\boldmath$\varepsilon$}},{\rm {\bf k}})$ is 
determined by the molecular transition and the unspecified initial 
configuration within band $S$. The second sum after the second equality was 
simplified using the unit sum of photon-number probabilities for each mode:
\begin{equation}
\label{eq9}
\sum\nolimits_{l=0}^\infty {\mathcal{P}_{{\mbox{\footnotesize \boldmath$\varepsilon$}},{\rm {\bf k}}} 
(l_{{\mbox{\footnotesize \boldmath$\varepsilon$}},{\rm {\bf k}}} )} =1.
\end{equation}
Assuming that the isotropic medium is linear and that the photon energy is $h\nu $, Eqs. [\ref{eq2a}] and [\ref{eq8}] combine to require that we \textit{define} the isotropic equilibrium $B$-coefficient spectrum for stimulated emission by averaging over all modes with frequency $\nu_{{\rm {\bf k}}} $ for an equilibrium initial configuration within band $S$:
\begin{equation}
\label{eq10}
\left\langle {{ }^{eq}\beta_{S\to R} ({\mbox{\boldmath$\varepsilon$}},{\rm {\bf 
k}})} \right\rangle_{{\mbox{\footnotesize \boldmath$\varepsilon$}},{\rm {\bf k}}} =b_{S\to R} 
(-\nu_{{\rm {\bf k}}} ,p,T)h\nu_{{\rm {\bf k}}} /V.
\end{equation}
With Eq. [\ref{eq10}], the first sum after the second equality in Eq. [\ref{eq8}] contains, in 
the limit of large volume, 
\begin{equation}
\label{eq11}
\sum\limits_{{\mbox{\footnotesize \boldmath$\varepsilon$}},{\rm {\bf k}},l} {\frac{\mathcal{P}_{{\mbox{\footnotesize \boldmath$\varepsilon$}},{\rm {\bf k}}} (l_{{\mbox{\footnotesize \boldmath$\varepsilon$}},{\rm {\bf k}}} 
)h\nu_{{\rm {\bf k}}} l_{{\mbox{\footnotesize \boldmath$\varepsilon$}},{\rm {\bf k}}} }{V}} 
=\int {u_{+}^{\nu } (\nu )d\nu } ,
\end{equation}
where $u_{+}^{\nu } $ is the average spectral density of electromagnetic energy per unit volume that appears in Eq. [\ref{eq2a}]. Similarly, the second sum after the second equality of Eq. [\ref{eq8}] contains
\begin{equation}
\label{eq12}
\sum\limits_{{\mbox{\footnotesize \boldmath$\varepsilon$}},{\rm {\bf k}}} {\frac{h\nu_{{\rm {\bf 
k}}} }{V}=\int {h\nu G^{\nu}_{+}(\nu ,p,T)d\nu } } ,
\end{equation}
where $G^{\nu}_{+}$ is the spectral density of electromagnetic modes per unit volume that appeared in the Planck blackbody radiation spectral density. Using Eqs. [\ref{eq8}]-[\ref{eq12}], the conditional transition probability per unit time for total emission becomes
\begin{align}
\begin{split}
\label{eq13}
 { }^{em}\Gamma_{S\to R} 
& =\sum\limits_{{\mbox{\footnotesize \boldmath$\varepsilon$}},{\rm {\bf 
k}},l} {b_{S\to R} (-\nu_{{\rm {\bf k}}} ,p,T)[\mathcal{P}_{{\mbox{\footnotesize \boldmath$\varepsilon$}},{\rm {\bf k}}}
 (l_{{\mbox{\footnotesize \boldmath$\varepsilon$}},{\rm {\bf k}}} )h\nu_{{\rm 
{\bf k}}} (l_{{\mbox{\footnotesize \boldmath$\varepsilon$}},{\rm {\bf k}}} +1)/V]} \\ 
& =\int_{{\kern 1pt}0}^{+\infty } {b_{S\to R} (-\nu ,p,T)[u_{+}^{\nu } (\nu 
)+h\nu G^{\nu }_{+}(\nu ,p,T)]d\nu } , \\ 
\end{split}
\end{align}
in which the integral of the first product after the second equality is the conditional transition probability per unit time for stimulated emission from $S$ to $R$ in Eq. [\ref{eq2a}] and the integral of the second product after the second equality is the conditional transition probability per unit time for spontaneous emission from $S$ to $R$ in Eq. [\ref{eq2b}]. This last identification proves the first generalized Einstein relation:
\begin{subequations}
\label{eq14}
\begin{equation}
\label{eq14a}
a_{S\to R}^{\nu } (-\nu ,p,T)=h\nu G^{\nu }_{+}(\nu ,p,T)b_{S\to R} (-\nu 
,p,T).
\end{equation}
For any given transition, the spectral density for spontaneous emission is equal to the product of the photon energy, the spectral density of electromagnetic modes per unit volume, and the Einstein $B$-coefficient spectrum for stimulated emission. As can be seen from Fig. \ref{fig:2}, Eq. [\ref{eq14a}] does not depend on assuming detailed balance or Planck blackbody radiation, it results directly from the electromagnetic mode density connection between spontaneous and stimulated emission.

With Eq. [\ref{eq14a}] proven, Eqs. [\ref{eq5}] and [\ref{eq6}] uniquely establish the generalized Einstein relation between the stimulated reverse transition from $S$ to $R$ and the stimulated forward transition from $R$ to $S$:
\begin{equation}
\label{eq14b}
b_{S\to R} (-\nu ,p,T)=\frac{{ }^{eq}\mathcal{P}_{R} (p,T)}{{ }^{eq}\mathcal{P}_{S} 
(p,T)}b_{R\to S} (\nu ,p,T)\exp (-h\nu /k_{\text{B}} T).
\end{equation}
\end{subequations}
Except in special circumstances, this single-molecule relationship does not necessarily hold between the average spectra of an inhomogeneous sample.\cite{RN2971} If all of $S$ lies energetically above all of $R$, then the forward transition from $R$ to $S$ is absorption and the reverse transition from $S$ to $R$ is stimulated emission. At thermodynamic equilibrium, the time-averaged, single-molecule results of Eq. [\ref{eq14}] are valid for any temperature above zero.

For interband transitions, Eq. [\ref{eq14}] provides five pairwise relationships between the four spectra in Eq. [\ref{eq1}]. Although $6=4\cdot 3/2$ pairwise relationships are possible among four spectra, there is no direct sixth relationship between the two $A$-coefficient spectral densities. In principle, either $B$-coefficient spectrum determines its $A$-coefficient spectral density and determines both reverse spectra up to a common constant multiplier $^{eq}\mathcal{P}_S/^{eq}\mathcal{P}_R$, so it determines all four lineshapes. (Unlike spectra, lineshapes such as $b_{S \to R}(\nu)/B_{S \to R}$ contain no information about transition strength.) Alternatively, if both $A$-coefficient spectral densities are non-zero, they can determine all four spectra.\cite{RN3956}  If one $A$-coefficient spectral density were zero, Eq. [\ref{eq14}] would provide three pairwise relationships among the three non-zero spectra, paralleling the three pairwise relationships among the three Einstein coefficients for line spectra. For a finite linewidth, the Einstein coefficients need not obey Einstein's relationships.

Intraband transitions with $R=S$ have only two Einstein coefficient spectra. For intraband transitions, Eq. [\ref{eq14a}] relates $a^{\nu}_{R \to R}(-\nu)$ to $b_{R \to R}(-\nu)\theta(\nu)$ and Eq. [\ref{eq14b}] constrains $b_{R \to R}(-\nu) = b_{R \to R}(\nu)\exp(-h\nu/k_\text{B} T)$ at equilibrium. As a result, any one half-spectrum determines all three non-zero half-spectra for intraband transitions.

\subsection*{Einstein's Special Case}
Einstein considered a stationary molecule that is isolated in vacuum and has infinitely narrow line transitions between idealized energy levels ($r$ and $s$) with temperature-independent quantum properties. It will be shown here that imposing Einstein's temperature-independent quantum properties and intramolecular Boltzmann distribution on {Eq.} [\ref{eq14}] gives the Bohr frequency condition and Einstein's relations for line spectra.

For a single isolated molecule, the intramolecular Boltzmann probability ratio for occupation of levels $s$ and $r$ is

$(^{eq}\mathcal{P}_{s}/^{eq}\mathcal{P}_{r})$ = ($g_{s}$/$g_{r}$)exp[-(${E}_{s}-{E}_{r}$)/$k_{\text{B}}{T}$],

\noindent where ${E}_{s}$ and $g_{s}$ are the quantum energy and degeneracy of level $s$. For transitions between energy levels of a single molecule in vacuum, {Eq.} [\ref{eq14b}] becomes
\begin{flalign}
\begin{split}
\label{eq23}
& g_{s} b_{s\to r} (-\nu ,T) \\
& =g_{r} b_{r\to s} (\nu ,T)\exp [-(h\nu -(E_{s}-E_{r}))/k_{\text{B}} T].
\end{split}
\end{flalign}

Einstein's derivation requires a single (as yet unspecified) frequency $\nu_{sr}$ for transitions between \textit{r} and \textit{s}, so that

$b_{r\to s} (\nu ,T) = B_{r\to s} \delta (\nu - \nu_{sr})$

\noindent and

$b_{s\to r} (-\nu ,T) = B_{s\to r} \delta (\nu - \nu_{sr})$.

\noindent  Following Einstein,\cite{RN978} we require that the quantum level degeneracies, \textit{B} coefficients, energy level difference, and transition frequency $\nu_{sr}$ are constants, independent of temperature.\cite{RN3951} Substituting the above spectra, integrating both sides of Eq. [\ref{eq23}], and using Eqs. [\ref{eq1c}] and  [\ref{eq1a}]  gives

$g_{s} B_{s\to r} =g_{r} B_{r\to s} \exp [-(h\nu_{sr} -(E_{s}-E_{r}))/k_{\text{B}} T].$

\noindent Since the quantum properties are all independent of temperature, we immediately obtain the Bohr frequency condition,

$\nu_{sr} =(E_{s} -E_{r} )/h,$

\noindent and Einstein's absorption-stimulated emission relation for line spectra in vacuum,

$g_{s} B_{s\to r} =g_{r} B_{r\to s}. $

Using the spectral density of modes from Eq. [\ref{eq7}] with vacuum refractive index $n = 1$, and substituting Eq. [\ref{eq14a}] into Eqs. [\ref{eq1a}] and [\ref{eq1b}], we see that it reduces to Einstein's spontaneous-stimulated emission relation

$A_{s\to r} =(8\pi h\nu_{sr}^{3} /c^{3})B_{s\to r} $.

\noindent As in Einstein's treatment,\cite{RN978} the Bohr transition frequency emerges as a consequence of an intramolecular Boltzmann distribution and his requirement that properties of idealized infinitely narrow quantum levels be constants, independent of temperature; conservation of energy was not directly invoked. None of the additional requirements or results of this subsection are used elsewhere in this paper except to discuss Einstein's special case in the subsection on Spectroscopic Thermodynamics.

\section*{Results}
\subsection*{Radiative Thermalization}
The generalized Einstein relations are a consequence of imposing equilibrium with Planck blackbody radiation and equilibrium band probabilities (which remain undetermined at this point) on hypothesized rate expressions for single-photon transitions. Following Einstein,\cite{RN978} if our hypothesis and assumptions are correct, the resulting molecule-radiation interaction must, all by itself, drive both radiatively coupled molecular degrees of freedom and radiation field to a dynamic equilibrium that agrees with the theory of heat. In this section, we show that the generalized Einstein relations have implications, beyond those directly mandated by hypothesis and assumptions, for molecular equilibrium within bands, molecular translation, and the photon number distribution.

The factor of $\exp (-h\nu /k_{\text{B}} T)$ on the right hand side of Eq. [\ref{eq14b}] red-shifts stimulated emission to lower frequencies than absorption. As illustrated in Fig. \ref{fig:1}d -- \ref{fig:1}f, this frequency shift becomes significant for linewidths that are appreciable compared to the thermal energy $k_{\text{B}} T$. This Stokes' shift between absorption and emission\cite{RN3252} was first qualitatively explained by Einstein as caused by thermal dissipation of excess molecular energy after excitation by one photon and before emission of another.\cite{RN3926} Here, we have found the quantitative form of the Stokes' shift that is required for equilibrium with Planck blackbody radiation. In particular, it holds for radiative line broadening in vacuum, where excitation by blackbody radiation and energy conservation \textit{directly} dictate the total thermal emission spectral density so that it differs from that produced by non-equilibrium resonance fluorescence with spectrally flat excitation\cite{RN1419}. This equilibrium result for purely radiative broadening does not require equilibration within the upper band before emission. As illustrated in Fig. \ref{fig:1}, this Stokes' shift is generated by different thermal equilibrium distributions within the initial band for absorption and emission transitions. A Stokes' shift between absorption and stimulated emission cross-sections of similar form (without negative frequencies and the signed cross sections to be used below) has been previously obtained from an equilibrium Boltzmann distribution for quantum level occupation probabilities within each band [see \cite{RN2267} and references cited in \cite{RN2971}]; this prior result is subject to two additional restrictions to be developed below. In contrast, we have not presumed anything about linewidth, molecular quantum statistics, or the equilibrium energy distribution within a band -- rather, equilibrium with Planck blackbody radiation generates, all by itself, a Stokes' shift that reflects equilibrium energy distributions within bands from the theory of heat. Equation [\ref{eq14b}] demonstrates that the equilibrated Stokes' shift always has the same form when written in terms of Einstein $B$-coefficient spectra.

Although no assumptions have been made about the molecular quantum statistics, Eqs. [\ref{eq2}],[\ref{eq6}] and [\ref{eq14}] predict the translational velocity probability distribution for a single molecule in field-free vacuum. Einstein proposed that a molecule in vacuum always directionally absorbs or emits a photon with momentum of magnitude $h\nu {\kern 1pt}/{\kern 1pt}c$ into a single mode.\cite{RN978} Einstein demonstrated that momentum-conserving, completely directional absorption and emission in a vacuum blackbody radiation field do not disturb the average translational kinetic energy of a Maxwell-Boltzmann velocity probability distribution when Doppler frequency shifts, transformation of the electromagnetic energy density, aberration, and the molecular photon recoil are taken into account (to first order in $v/c$) in the molecular rest frame.\cite{RN978} Einstein treated molecular translation with non-relativistic classical mechanics and found that the average linear dissipative drag from net absorption steadily damps the molecular velocity, but that random fluctuations in velocity from photon recoil counterbalance the drag to sustain equilibrium. There are no essential difficulties in using Einstein coefficient spectra to adapt Einstein's demonstration.\cite{RN23} One can then adapt Milne's completion of Einstein's treatment\cite{RN3907} to calculate how vacuum blackbody radiation drives any non-equilibrium velocity distribution to the thermal equilibrium Maxwell-Boltzmann probability distribution. Finally, Einstein's theory of Brownian motion\cite{RN985} can be used to calculate how vacuum blackbody radiation drives any non-equilibrium molecular spatial distribution to the spatially uniform equilibrium probability distribution. This demonstrates (to first order in $v/c$) that Einstein coefficient spectra in the molecular rest frame combine with vacuum blackbody radiation to drive translational equilibrium in the rest frame where blackbody radiation is isotropic.\cite{RN3924,RN3908} Einstein's demonstration was based on an intramolecular Boltzmann distribution for quantum level occupation probabilities.\cite{RN978} In contrast, the only statistical assumption that entered the derivation here was equilibrium with homogeneous and isotropic Planck blackbody radiation, but the vacuum Planck law requires infinitely dilute molecules in free space, so the quantum statistical consequences of that requirement appear here. 

Bothe\cite{RN432} used Einstein coefficients for line spectra to obtain the Bose photon number distribution\cite{RN426} for the blackbody radiation field by treating all emission as a single process and requiring, for each frequency and each $l$, detailed-balance equality between the equilibrium total rate for absorption from all modes with ($l + 1$) photons and the equilibrium total rate for emission from all modes with $l$ photons. There are no essential difficulties with using Einstein coefficient spectra and the generalized Einstein relations in Bothe's argument,\cite{RN3927} which is based on a stronger form of detailed balance than Eq. [\ref{eq4}]. Again, no assumptions about molecular quantum statistics are needed.

\subsection*{Transition Cross Sections}
Einstein coefficient spectra directly obey detailed balance and the generalized Einstein relations, but are asymmetric with respect to zero frequency so that their relationship to causality is an indirect one through transition cross section spectra. If the molecules in an ensemble are isotropically oriented on average and absorb independently of each other,\cite{RN46} the Einstein $B$-coefficient spectra for isotropic and unpolarized light can be used to calculate net absorption from a polarized beam of light. If, in addition, the solution is homogeneous and uniform in the electromagnetic sense,\cite{RN3928} the beam of light is normally incident, and the beam of light is so weak that band populations and distributions practically remain at equilibrium, then the Beer-Lambert law holds. For a beam of light, the steady-state spectral irradiance is
\begin{equation}
\label{eq15}
I_{+}^{\nu } (\nu ,z)=u_{+}^{\nu } (\nu ,z){\kern 1pt}\cdot v_{1} (\nu 
,p,T),
\end{equation}
where the energy transport velocity $v_{1} $ reduces to the group velocity 
$v_{g} (\nu ,p,T)=c/[\partial(\nu n(\nu ))/{\kern 1pt}\partial \nu ]$ in weakly dispersive transparent media.\cite{RN477} Taking into account loss and gain in the electromagnetic equation of continuity, propagation of the spectral irradiance with sample depth $z$ involves a sum over all transitions from all initial bands(\ref{eq11}) 
\begin{equation}
\label{eq16}
\frac{I_{+}^{\nu } (\nu ,z)}{I_{+}^{\nu } (\nu ,0)}=\exp \left[ 
{-\sum\limits_{R,S} {N_{R} \sigma_{R\to S} } (\nu ,p,T)z} \right],
\end{equation}
where $N_{R}$ is the molecular number density in initial band $R$, and each transition cross section spectrum is
\begin{flalign}
\begin{split}
\label{eq17}
& \sigma_{R\to S} (\nu ,p,T) \\
& ={h\nu [b_{R\to S} (\nu ,p,T)-b_{R\to S} (-\nu 
,p,T)]} \mathord{\left/ {\vphantom {{h\nu [b_{R\to S} (\nu ,p,T)-b_{R\to S} 
(-\nu ,p,T)]} {{\kern 1pt}v_{1} (\nu ,p,T)}}} \right. 
\kern-\nulldelimiterspace} {{\kern 1pt}v_{1} (\nu ,p,T)}.
\end{split}
\end{flalign}
For positive frequency $\nu $, $b(\nu )$ represents absorption and $b$(-$\nu )$ represents stimulated emission. The transition cross section is positive for net absorption and negative for net stimulated emission. In contrast to prior work, each transition cross section includes the opposite effects of photon number losses from absorption and gains from stimulated emission that both originate from a single $B$-coefficient spectrum. The sum in Eq. [\ref{eq16}] includes intraband transitions with $R = S$. Compared to the usual formula in which all cross sections are defined as positive,\cite{RN3162} Eqs. [\ref{eq16}] and [\ref{eq17}] algebraically distinguish absorption from stimulated emission through the sign of the transition cross section.

\begin{figure}[t!]
\centering
\includegraphics[width=.8\linewidth]{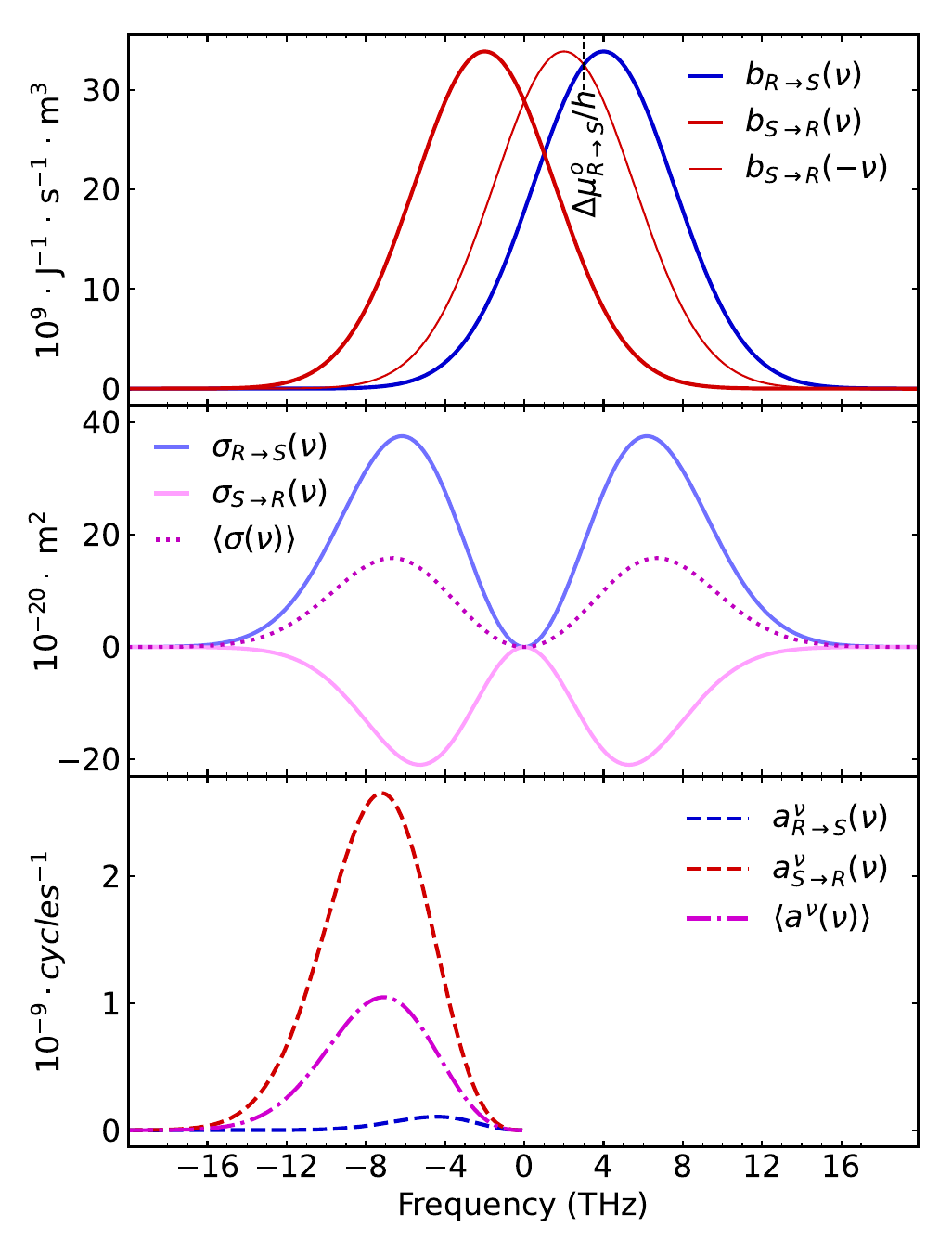}
\caption{Low-frequency transitions between energetically 
overlapping bands. Top) Einstein $B$-coefficient spectra for $R$ to $S$ (thick blue) and $S$ to $R$ (thick red) transitions. All spectra shown are completely specified by a Gaussian lineshape for the mostly absorptive $R$ to $S$ transition centered at $\nu_{R\to S} =$ 4 THz, a $B$ coefficient
$B_{R\to S} =e^{2}/(4\epsilon_{0} m_{e} h\nu_{R\to S} )\approx $ 3.0\texttimes10$^{23}$ m$^{3}$/J$\cdot$s$^{2}$,
\cite{RN32} 
an equilibrium population ratio specified by a change in standard chemical potential of $\Delta \mu_{R\to S}^{\text{o}} /h =$ 3 THz, a Stokes' shift of $(2\lambda ) =$ 2 THz and a temperature $T$ $=$ 300K ($k_{\text{B}} T/h\approx $ 6.2 THz). Positive transition frequencies indicate absorption and negative transition frequencies indicate stimulated emission. The thin red curve is the frequency-reversed $B$-coefficient spectrum of the $S$ to $R$ transition.
Middle) The $R$ to $S$ (light blue) and $S$ to $R$ (pink) transition cross-sections between the two energetically overlapping bands. Each has both absorption and stimulated emission contributions that cancel at zero frequency and partially cancel nearby. A positive transition cross section represents net absorption and a negative transition cross section represents net stimulated emission. Transition cross sections are functions of the radiation frequency. The thermally averaged transition cross-section (dotted magenta) weights each transition cross section by the Boltzmann population of the initial band. For simplicity, $R$ and $S$ are the only bands and the refractive index is $n$ $=$ 1. Bottom) Einstein $A$-coefficient spontaneous emission spectral densities for the $R$ to $S$ transition (dashed blue) and $S$ to $R$ transition (dashed red) as functions of the transition frequency. The initial band population weighted average (dot-dashed magenta) is proportional to the spontaneous emission photon number spectral density.}
\label{fig:3}
\end{figure}

Figure \ref{fig:3} shows a pair of $B$-coefficient spectra that obey the generalized Einstein relation and the transition cross section for each. For transitions between a pair of bands, if one $B$-coefficient spectrum is confined to positive transition frequencies and the other $B$-coefficient spectrum is confined to negative transition frequencies, then McCumber's broadband relations between absorption and stimulated emission cross sections\cite{RN2267} become accurate for the homogeneously broadened laser transitions that motivated his pioneering work. This requires linewidths that are narrow compared to the average photon energy in both absorption and emission (the "if" condition above requires a second necessary restriction on the linewidth to be introduced in the subsection on Extreme Stokes' Shifts). The forward and reverse transition cross sections in Fig. \ref{fig:3} illustrate a departure from McCumber's relation outside its limit of validity.

Up to this point, negative transition frequencies have referred to stimulated emission and the frequency of the electromagnetic field has always been positive. Equations [\ref{eq15}] - [\ref{eq17}] are equally valid for positive and negative electromagnetic field frequency $\nu $. From this point on, this allows us to adopt the complex-valued exponential Fourier transform view in which positive and negative frequencies (rather than sines and cosines) are needed to form a complete basis for the electromagnetic fields. Equations [\ref{eq15}] - [\ref{eq17}] have the same form for a positive-frequency spectral irradiance or for a spectral irradiance that is an even function of frequency over the entire real axis. Since the refractive index, and hence the energy transport velocity, is an even function of frequency, the transition cross sections are defined over the entire real frequency axis as even functions of frequency: $\sigma_{R\to S} (-\nu ,p,T)=\sigma_{R\to S} (\nu ,p,T)$.

When considered over the entire frequency axis, the physical interpretation of the two $B$-coefficient spectra in Eq. [\ref{eq17}] must be expanded. For negative values of the frequency, $b(\nu )$ represents stimulated emission and $b$(-$\nu )$ represents absorption, so each changes its nature upon crossing zero frequency. With this expanded physical interpretation, the two terms in Eq. [\ref{eq17}] parallel the positive and negative frequency terms in the exact rotating wave decomposition of the impulse response and susceptibility for interband transitions.\cite{RN90,RN3946} In both decompositions, terms that cross zero frequency indicate that photons can be both absorbed and emitted in the same molecular transition direction.\cite{RN713} Wiersma and co-workers have shown that neglecting rotating wave decomposition terms that cross zero frequency by making the rotating wave approximation generates errors for a model of the broad visible absorption spectrum of the solvated electron in water,\cite{RN213} so both terms in Eq. [\ref{eq17}] can be simultaneously important for optical transitions.

We now consider low-frequency behavior and detailed balance for transition cross sections. Each transition cross section in Eq. [\ref{eq17}] is even, so because each $B$-coefficient spectrum is finite for a finite-sized molecule, each transition cross section has a lowest order frequency variation as the square of the frequency (or a higher even power) near zero frequency. This lowest-order variation with the square of the frequency is an experimentally verified aspect of the Van Vleck-Weisskopf impact theory for the pressure-broadened net absorption lineshape in gases.\cite{RN3500,RN3414,RN327,RN3930} Although the Van Vleck-Weisskopf lineshapes are the only lineshapes justified by a microscopic theory that have been proven to satisfy detailed balance between absorption and emission, this agreement is restricted to low-frequency classical Rayleigh-Jeans blackbody radiation.\cite{RN3499} In contrast, Eqs. [\ref{eq4}], [\ref{eq6}], [\ref{eq14}], [\ref{eq15}], and [\ref{eq17}] show that the cross-sections found here obey detailed balance with Planck blackbody radiation:

\begin{align}
\begin{split}
\label{eq18}
& h\nu [{ }^{eq}\mathcal{P}_{R} a_{R\to S}^{\nu } (-\nu ,p,T)+{ }^{eq}\mathcal{P}_{S} a_{S\to 
R}^{\nu } (-\nu ,p,T)] \\ 
& \quad 
 =[{ }^{eq}\mathcal{P}_{R} \sigma_{R\to S} (\nu ,p,T)+{ }^{eq}\mathcal{P}_{S} 
\sigma_{S\to R} (\nu ,p,T)] \\
 & \quad \quad \cdot u_{\text{BB+}}^{\nu } (\nu ,p,T)v_{1} (\nu ,p,T).
\end{split}
\end{align}

For each pair of bands, the equilibrium statistical average of the spontaneously emitted power is equal to the equilibrium statistical average rate at which energy is absorbed (absorption minus stimulated emission) for every frequency.\cite{RN3929} Summing Eq. [\ref{eq18}] over all bands gives van Roosbroeck and Shockley's less specific relationship\cite{RN3495} between the total rates. The quantities in brackets can be compared in Fig. \ref{fig:3}. Equation [\ref{eq18}] shows that the fourth Einstein coefficient spectrum introduced here is necessary to reconcile the classical and quantum frequency regimes. This more symmetrical relationship between statistical average spontaneous emission and statistical average absorption differs practically from all prior work. Figure \ref{fig:3} shows the statistical average spontaneous emission and the forward and reverse transition cross sections with widths comparable to both the photon energy and the thermal energy; these do not obey the relationships in ref. \cite{RN3500,RN3499,RN2267}. Such transitions occur in the frequency range useful for thermal imaging,\cite{RN3539} where band populations are quite sensitive to temperature changes.

Causality imposes global requirements [dispersion relations\cite{RN2558}] on each frequency-dependent transition cross-section and its associated frequency-dependent phase shift so that transmitted signals cannot precede speed-of-light propagation of their inputs in the time domain. The transition cross sections have even frequency-domain symmetry and are continuous with a continuous first derivative, as expected for a finite system of bound charges. Their compatibility with causality thus follows from Titchmarsh's theorem\cite{RN2558} and the observation that the transition cross sections are square integrable. The phase shift spectrum associated with a transition cross section spectrum can be calculated by Kronig's method.\cite{RN90,RN1912}

\subsection*{Spectroscopic Thermodynamics}
We can obtain powerful additional results by allowing the sample to contain many identical molecules that obey Maxwell-Boltzmann statistics and recognizing that the ratio ${ }^{eq}\mathcal{P}_{R} /{ }^{eq}\mathcal{P}_{S} $ is an equilibrium constant. For simplicity, we start by assuming that the molecules in band $S$ behave as an ideal chemical constituent (ideal gas, ideal mixture,\cite{RN1986} 
ideal solution,\cite{RN1791} \textit{etc.}) so that
\begin{equation}
\label{eq19}
\frac{{ }^{eq}\mathcal{P}_{S} (p,T)}{{ }^{eq}\mathcal{P}_{R} (p,T)}=K_{R\to S} (p,T)=\exp 
[-\Delta \mu_{R\to S}^{\text{o}} (p,T)/k_{\text{B}} T],
\end{equation}
where $K_{R\to S} $ is the thermodynamic equilibrium constant for the thermal equilibrium reaction $R\to S$, 
\begin{equation}
\label{eq20}
\Delta \mu_{R\to S}^{\text{o}} (p,T)=\mu_{S}^{\text{o}} (p,T)-\mu 
_{R}^{\text{o}} (p,T)
\end{equation}
is the change in standard chemical potential for $R\to S$, and $\mu _{S}^{\text{o}} (p,T)$ is the standard chemical potential\cite{RN2661} for a molecule in band $S$. The standard chemical potential is the per-molecule form of the standard Gibbs free energy, $G_{S}^{\text{o}} =N_{\text{A}} \mu _{S}^{\text{o}} $, where $N_{\text{A}} $ is Avogadro's number. It is an intrinsic material property, independent of molecular number density. In contrast, the chemical potential depends on $N_{S} $, the number density in band $S$, the standard chemical potential, and the standard number density $N^{\text{o}}$ as $\mu_{S} (N_{S} ,p,T)=\mu_{S}^{\text{o}} (p,T)+k_{\text{B}} T\ln (N_{S} /N^{\text{o}})$.\cite{RN1791,RN2661} Since this is a unimolecular reaction, the change in standard chemical potential is independent of the chosen standard states.

There is no essential difficulty in generalizing to non-ideal thermodynamic constituents or the presence of external fields -- then all quantities in this section depend on the mole fractions $\{x_{i} \}$ of the minimum number of chemical components necessary to specify system composition\cite{RN1791} plus any external fields, the thermodynamic equilibrium constant $K_{R\to S} (p,T)$ is replaced by a composition and field dependent number density equilibrium constant $K_{R\to S}^{N} (p,T,...\;\,x_{i} ,...)={ }^{eq}N_{S} /{ }^{eq}N_{R} $, and the standard chemical potentials $\mu_{S}^{\text{o}} (p,T)$ are replaced by composition and field dependent formal chemical potentials $\mu_{S}^{\text{o$'$}} (p,T,...\;\,x_{i} ,...)$. The term formal chemical potential is used by analogy to non-ideal formal electrode potentials,\cite{RN235} which replace standard electrode potentials for specified non-standard conditions.

With Eq. [\ref{eq19}], the ideal Boltzmann form of the generalized Einstein relation in Eq. [\ref{eq14b}] becomes
\begin{flalign}
\begin{split}
\label{eq21}
& b_{S\to R} (-\nu ,p,T) \\
& =b_{R\to S} (\nu ,p,T)\exp [-(h\nu -\Delta \mu_{R\to S}^{\text{o}} (p,T))/k_{\text{B}} T].
\end{split}
\end{flalign}
The frequency-dependent $\exp (-h\nu /k_{\text{B}} T)$ factor relates the lineshapes of forward and reverse spectra; the frequency-independent change in standard chemical $\Delta \mu_{R\to S}^{\text{o}} $ relates their magnitudes. The forward and reverse $B$-coefficient spectra are equal at the photon energy equal to the change in standard chemical potential (as shown in the top panel of Fig. \ref{fig:3}).\cite{RN3953} According to Bohr's interpretation of the Rydberg-Ritz combination principle, a small number of quantum energy levels determines the frequencies for the larger number of spectroscopic transitions between those levels.\cite{RN3663} Since the standard chemical potential is a thermodynamic state function, we assert here that standard chemical potentials of bands will also obey the Rydberg-Ritz combination principle.

To make contact with Einstein's special case of line spectra, we consider an idealized quantum level $s$ of an isolated and stationary molecule in vacuum, where the standard chemical potential becomes a function of temperature alone,
\begin{equation}
\label{eq22}
\mu_{s}^{\text{o}} (T)=E_{s} -k_{\text{B}} T\ln (g_{s} )+constant,
\end{equation}
$E_{s}$ is the energy of the quantum level, $g_{s}$ is its degeneracy, and the constant is needed to put different molecules on the same scale of standard chemical potentials. Substituting Eq. [\ref{eq22}] into Eq. [\ref{eq20}] and [\ref{eq21}] gives Eq. [\ref{eq23}]. In the generalized Einstein relation of Eq. [\ref{eq21}], the entropic contribution to the standard chemical potential generalizes the degeneracy of a quantum level.

\subsection*{Extreme Stokes' Shifts}

\begin{figure*}
\centering
\includegraphics{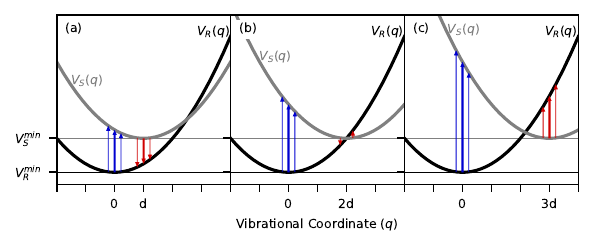}
\caption{Transitions between molecular bands with displaced potential curves. The vibrational potential energy curves for electronic states $R$ and $S$ have the same harmonic force constant and same energy difference between potential minima throughout. Panels a), b), and c) show increasing displacements of the equilibrium vibrational coordinate for band $S$. Forward (absorption) transitions from the thermal equilibrium coordinate distribution on $R$ to $S$ are represented by blue arrows, and reverse transitions from the equilibrium coordinate distribution on $S$ to $R$ are represented by red arrows. Transition linewidths and the Stokes' shift both increase as the vibrational displacement increases from left to right. The Stokes' shifted reverse transition is predominantly emissive in a), centered at zero frequency in b), and predominantly absorptive in c).}
\label{fig:4}
\end{figure*}

The generalized Einstein relation in Eq. [\ref{eq14b}] has the thought-provoking consequence that sufficiently broad forward transitions that are practically absorptive can generate a Stokes' shift so large that the reverse transition becomes practically absorptive instead of emissive. Figure \ref{fig:4} illustrates how a standard adiabatic model for transitions between electronic states with vibrationally displaced potential energy curves\cite{RN723} can give such results. In this model, the photon energy for a transition is equal to the vertical energy difference between potential energy curves, so that a thermal distribution of vibrational coordinates on the lower electronic curve broadens the electronic transition.  This vibrational broadening gives rise to a progressively broader and higher frequency absorption spectrum as the upper electronic curve is displaced to the right without any change in the minimum-to-minimum energy difference. As the (forward) absorption spectrum shifts to higher frequencies, the (reverse) emission spectrum shifts to lower frequencies; the reverse transition hits zero frequency in Fig. \ref{fig:4}b, where the two curves cross at the minimum of the upper curve; the reverse transition is absorptive in Fig. \ref{fig:4}c because the upper curve is displaced so far to the right that its minimum lies outside and below the lower curve. These behaviors do not conflict with the generalized Einstein relations because the generalized Einstein relations connect absorption in the forward transition to total emission at the same frequency in the reverse transition. This simplified model provides an example in which a generalization of Einstein's relations must allow extremely large Stokes' shifts to generate absorption instead of emission. The reversal from emissive to absorptive character for a large Stokes' shift arises naturally by using signed frequencies in a set of 4 Einstein coefficient spectra.

These consequences of the generalized Einstein relations developed here are most readily illustrated mathematically by using a Gaussian lineshape for the forward $B$-coefficient spectrum from $R$ to $S$. For this lineshape, Eq. [\ref{eq14b}] shows that the $B$-coefficient spectrum for the reverse transition from $S$ to $R$ must also be a Gaussian with the same variance and that the variance $\Delta ^{2}$ and Stokes' shift $(2\lambda )$ between absorption and emission are related by the thermal energy: $\Delta^{2}=(2\lambda )k_{\text{B}} T/h$.\cite{RN2971} (The same equation has previously been obtained from semi-classical displaced harmonic oscillator models for electronic transitions,\cite{RN3945} but the generalized Einstein relations show that any homogeneous Gaussian lineshape has this Stokes' shift.) If the center frequency for the forward transition is $\nu_{R\to S} $, then the dimensionless parameter $\Delta^{2}/(\nu_{R\to S} k_{\text{B}} T/h)$ becomes one when $(2\lambda )=\nu_{R\to S} $ so that the reverse transition is centered at $\nu =0$ (as in Fig. \ref{fig:4}b). With such an improbably precise coincidence, the cross section for the reverse transition would vanish identically for all frequencies, $\sigma_{S\to R} (\nu )=0$, yet its spontaneous emission spectral density $a_{S\to R}^{\nu } (-\nu )$ would remain non-zero. For larger linewidths, $b_{S\to R} (-\nu ,p,T)$ becomes centered at negative $\nu $, so that $b_{S\to R} (\nu ,p,T)$ is centered at positive $\nu $ and the reverse transition from $S$ to $R$ becomes absorptive (as in Fig. \ref{fig:4}c). A large homogeneous absorption linewidth can generate a Stokes' shift so extreme that the reverse transition becomes predominantly absorptive. Furthermore, so long as $\Delta /(k_{\text{B}} T/h)$ is sufficiently large, such an extreme Stokes' shift can occur for small $\Delta /\nu_{R\to S} $.

As mentioned previously, prior detailed-balance relations that account for a Stokes' shift\cite{RN2267,RN2971} are practically limited to linewidths much narrower than the transition frequency, $\Delta \ll \nu_{R\to S} $. For a homogeneous Gaussian lineshape, the above paragraph places a second necessary restriction on the variance, $\Delta^{2}\ll \nu_{R\to S} k_{\text{B}} T/h$, which depends on both the transition frequency and the thermal energy. These two restrictions are independent. When either of these two conditions is not satisfied, the theory with 4 Einstein coefficient spectra, signed frequencies, and signed transition cross sections developed here allows continuous changes from emissive to absorptive transitions with dramatic consequences.

\section*{Discussion}
The generalized Einstein relations between absorption and emission spectra are exact at thermodynamic equilibrium. However, there is essentially no thermal equilibrium emission from excited electronic bands at room temperature, so electronic emission spectra are measured by non-equilibrium luminescence, incandescence, or stimulated emission. The derivation of Eq. [\ref{eq14a}] connecting spontaneous and stimulated emission actually requires only that their rates be well defined. Application of the Einstein line spectra $A$ and $B$ relationships to luminescence and absorption line spectra implicitly supposes rapid equilibrium among the degenerate states within a quantum level. Similarly, equilibrium Einstein coefficient spectra become applicable to non-equilibrium luminescence after equilibrium within the luminescent band (but not between different bands), which establishes a non-equilibrium chemical potential for the luminescent band. This circumstance is called thermal quasi-equilibrium. In thermal quasi-equilibrium, the conditional probabilities for configurations within each band take their equilibrium values, but the prior probability for the band, $\mathcal{P}_{S} $, deviates from its equilibrium probability ${ }^{eq}\mathcal{P}_{S} $. For a large molecule in a room temperature solution, a large body of evidence indicates that thermal quasi-equilibrium within excited electronic states is usually established on a few picosecond timescale.\cite{RN2008,RN208} Picosecond thermal quasi-equilibrium is even more firmly established within the conduction and valence bands of semiconductors, where it is the criterion for the existence of quasi-Fermi levels.\cite{RN1795}

Non-equilibrium applications require that each band reach thermal quasi-equilibrium much faster than relaxation between separate bands, so that each band can be treated as a metastable thermodynamic constituent. For example, transitions between the upper and lower Dirac cones of graphene\cite{RN3957} should be treated as intraband transitions within a single band consisting of the double cone. If bands are in rapid equilibrium on the timescale of a slower measurement, it can sometimes be necessary or convenient to regard them as a single band. Conversely, bands originating from different components necessary to specify thermodynamic composition cannot be combined in this way. Steady-state luminescence weights spectra by quantum yield rather than radiative rate, so luminescence spectra are not necessarily proportional to the spectral density of the radiative rate that appears in Eq. [\ref{eq14a}]; this provides more opportunities for detecting ensemble inhomogeneity by comparing absorption and luminescence\cite{RN2971} than those implied by Eq. [\ref{eq14b}].

The generalized Einstein relations allow a broad absorption transition to generate such an extreme Stokes' shift that its reverse transition crosses zero frequency to become mainly absorptive instead of emissive. We suggest that one-electron intervalence-transfer absorption transitions in symmetrical mixed-valence complexes \cite{RN3945,RN3947,RN3954}  can be regarded as prototypical examples with $V_{S}^{\min } =V_{R}^{\min } $ in Fig. \ref{fig:4} and $\Delta \mu _{R\to S}^{\text{o}} =0$ by symmetry. In these transitions, an asymmetrically localized charge is transferred between two equivalent centers that are weakly coupled through an insulating bridge.
In fact, within the approximation of a Gaussian absorption lineshape, these intervalence-transfer absorptions obey $\nu_{R\to S } \approx \Delta^{2}/(2k_{\text{B}} T/h)$.\cite{RN3945, RN3954} In the context of the generalized Einstein relations, this known relationship between their center frequency and homogeneous linewidth arises from an extreme Stokes' shift of $(2\lambda )=2\nu_{R\to S} $ between forward and reverse absorption transitions with the same $B$-coefficient spectra. Such transitions can have visible absorption linewidths that are much less than the center photon energy but much greater than the thermal energy, with practically no emission\cite{RN3945} - the two charge configurations $R$ and $S$ can equilibrate through normal electron transfer.\cite{RN2212,RN3948} (If the electron-transfer coupling expands the coherent molecule-environment evolution beyond one final state during the radiative transition, then it modifies the spectra of both states and the two electronic states must both belong to the same band even at equilibrium.) The generalized Einstein relations still hold at equilibrium and the rate of spontaneous emission still balances the rate of net absorption at each frequency in Eq. [\ref{eq18}], but the equal equilibrium populations of the initial and final bands imply that the equilibrium conditional transition probability for spontaneous emission per unit time (${ }^{a}\Gamma_{S\to R} (p,T)$ in Eq. [\ref{eq2b}]) can be \textit{many orders of magnitude smaller} in relation to the integrated absorption cross section than for an ordinary forward-absorptive/reverse-emissive transition. In the asymmetrical circumstances of Fig. \ref{fig:4}c, the upper state still practically loses its equilibrium emission, but can decay by normal back-electron transfer, quenching, etc. Use of the generalized Einstein relations to identify Stokes' shifted absorption in other systems could provide insights into their spectroscopy, thermodynamics, reaction dynamics, and quasi-equilibrium radiative processes.

For interband transitions, the signed frequency and fourth Einstein coefficient introduced here can be practically important if either transition cross section has a low-frequency $\nu^{2}$ component [as for intraband transitions \cite{RN3500,RN3499,RN3414,RN327}], if $a^{\nu}_{S \to R}(-\nu) / \nu^{3}$ does not vanish in the low-frequency limit [as may be the case for the non-equilibrium spontaneous emission spectral density of the solvated electron – see ref. \cite{RN3940}], if the rotating wave approximation breaks down [as in the absorption spectrum of the solvated electron – see ref. \cite{RN213}], or if a Gaussian linewidth has variance $\Delta^{2}$ approaching or exceeding $(\nu_{R\to S} k_{\text{B}} T/h)$ [as in intervalence transitions \cite{RN3945,RN3947,RN3954} and the visible absorption spectrum of the solvated electron \cite{RN3939}]. Such matters probe the extreme wings of the lineshape, about which little is known, so the above diagnostic list may not be exhaustive. The practical need for a signed frequency and extra $A$ coefficient do not necessarily go hand-in-hand; for example, symmetrical one-electron intervalence transfer absorptions require a signed frequency and two $B$-coefficient spectra, but both $A$ coefficients are practically negligible. The signed frequency and extra $A$ coefficient might not be practically necessary for transitions in which the absorption cross section is linked to emission \cite{RN3955} by a generalized Einstein relation within measurement accuracy and $a^{\nu}_{S \to R}(-\nu) / \nu^{3}$ vanishes in the low-frequency limit within measurement accuracy.

The Einstein coefficient spectra and relations have been presented so as to illustrate their broad validity and how they can be extended. For instance, the spectral density of electromagnetic modes can be modified for an absorbing medium\cite{RN3920,RN2805} or a cavity\cite{RN3414} so long as the molecule-field coupling remains weak. Finally, the results developed here can be applied to other thermal excitation and de-excitation mechanisms involving absorption and emission of single quasi-particles (for example, treating phonon absorption and emission involves a different mode density and group velocity, as in Brillouin's discussion\cite{RN475} of generalizing the thermal radiation law of Balfour Stewart and Kirchhoff\cite{RN3663} to phonons).

\section*{Conclusions}
By exploiting quantum aspects of light, we have developed a picture of single-photon transitions between broadened molecular bands that can be treated as metastable constituent forms of a molecule within classical thermodynamics. The generalized Einstein relations presented here do not depend on molecular quantum or statistical mechanics. Rather, they establish temperature-dependent detailed-balance relationships between spectra that have both the non-specific character and the broad applicability of thermodynamic results. For a pair of levels, Einstein's theory has three independent parameters: one $B$ coefficient, one degeneracy ratio, and one Bohr transition frequency that combine to determine the line spectra. In parallel, the generalized Einstein relations have one $B$ coefficient, one change in standard chemical potential, and one underlying $B$-coefficient lineshape that combine to determine four different Einstein coefficient spectra between two bands. The generalized Einstein relations provide five pairwise relationships among the four Einstein coefficient spectra. Importantly, the generalized Einstein relations predict stimulated reverse lineshapes from stimulated forward lineshapes and \textit{vice versa}. In ordinary cases, where forward and reverse are absorptive and emissive, the general forward-reverse lineshape relation quantifies the Stokes' shift between absorption and emission that is always required by the theory of heat. In extreme cases, the Stokes' shift can be so large that the cross sections for both forward and reverse stimulated transitions become practically absorptive while extraordinarily slow spontaneous emission maintains detailed balance.

In conclusion, the generalized Einstein relations treat transitions between broadened metastable bands that have thermodynamic formal chemical potentials rather than transitions between sharp quantum levels that have energies and degeneracies. The relationships apply rigorously to thermal emission, and are expected to apply with high accuracy to emission from any band that has reached internal thermal quasi-equilibrium. This enables measurement of the intrinsic thermodynamic properties of thermalized excited states on ultrafast timescales. Such measurements could replace order of magnitude approximations for excited state equilibrium constants [developed by F\"{o}rster for excited state proton transfer,\cite{RN1108} by Marcus for excited state electron transfer,\cite{RN2212} and by others for specific photochemical reactions\cite{RN1277,RN2863}] with exact thermodynamic cycles that have spectroscopic accuracy. The determinations of the standard chemical potential for bright and dark excitons by Ryu \textit{et al}.\cite{RN2971} show that the generalized Einstein relations can also be used to measure non-equilibrium free energy in at least some circumstances. For a single molecule, detailed balance, the density of modes connection between spontaneous and stimulated emission, and Planck blackbody radiation dictate a Stokes' shift for emission, a Maxwell-Boltzmann translational velocity distribution, and relationships between Einstein coefficient spectra that are compatible with the uncertainty principle and encompass Einstein's results. The resulting relationships between transition cross sections are practically different from prior detailed-balance results\cite{RN3500,RN3499,RN2267} in the low frequency range useful for thermal imaging and can be dramatically different at any frequency for transitions with linewidths that exceed the thermal energy.

\section*{Acknowledgments}
We thank Aman Agrawal, Andres Montoya-Castillo, Niels Damrauer, Callum Douglass, Jennifer Sormberger, Richard Zare and especially Casey Hynes for helpful comments on the presentation of these results. This material is based upon work of J.R. and D.M.J. supported by the Air Force Office of Scientific Research under AFOSR award no. FA9550-18-1-0211 and upon work of S.Y. and D.M.J. supported by the National Science Foundation under award number CHE-2155010.

\bibliography{EinsteinREFv2}

\begin{thebibliography}{74}
\bibitem [{RN3({\natexlab{a}})}]{RN3938}%
  \BibitemOpen
  \bibinfo {note} {{IUPAC} allows a "molecular entity" to specify the molecule plus structural and/or quantum characteristics that are conserved on the timescale of an experiment. {The} calculation of total absorption from a sum over bands requires that each quantum state or thermodynamic molecular entity belong to one and only one band.}\BibitemShut {Stop}%
\bibitem [{RN3({\natexlab{b}})}]{RN3931}%
  \BibitemOpen
  \bibinfo {note} {A spectral density is characterized by invariant integrals over corresponding variable ranges and thus transforms under the change of variables theorem as $a^{\nu }(\nu ) = a^{\omega }(\omega =$2$\pi \nu )(d\omega $/$d\nu )$. {In} contrast, a spectrum is invariant with $b(\nu ) = b(\omega =$2$\pi \nu )$.}\BibitemShut {Stop}%
\bibitem [{RN3({\natexlab{c}})}]{RN3951}%
  \BibitemOpen
  \bibinfo {note} {Einstein explicitly stated that the statistical weights (called level degeneracies here) are numbers, independent of temperature, and that the ${A}$ and ${B}$ coefficients are constants. {We} think that {Einstein's} derivation uses an implicit assumption that the more clearly quantum mechanical ${A}$ and ${B}$ coefficients, energy-level difference, and level-to-level transition frequency are independent of temperature.}\BibitemShut {Stop}%
\bibitem [{RN3({\natexlab{d}})}]{RN3937}%
  \BibitemOpen
  \bibinfo {note} {For transitions between quantum states, complex-conjugate {Hermitian} matrix elements can be used to directly connect rate constants for absorption from and emission into a single mode without requiring time-reversal invariance. {This} approach is used in ref. \cite{RN2267}. {However}, these single-mode rate constant relationships from {Hermitian} matrix elements do not guarantee spatial equilibrium - see note \cite{RN3924} below.}\BibitemShut {Stop}%
\bibitem [{RN2()}]{RN23}%
  \BibitemOpen
  \bibinfo {note} {Using the correspondences (1/${S})p_{n}$exp(-${E}_{n}$/$k_{B}{T})$ to $^{eq}\mathcal{P}_{N}$, ${B}_{n}^{m}$ to $b_{N \to M}(\nu ,p,{T})$, ${B}_{m}^{n}$ to $b_{M \to N}(-\nu ,p,{T})$, and $p_{n}{B}_{n}^{m} = p_{m}{B}_{m}^{n}$ to {Eq.} [\ref{eq14b}], {Einstein's} demonstration carries over to broadened generalized {Einstein} coefficient spectra with only one slight modification. {The} modification is that the number of emission processes is not necessarily exactly equal to the number of absorption processes, but only equal on average (which is also the situation in a multi-level system).}\BibitemShut {Stop}%
\bibitem [{RN4()}]{RN46}%
  \BibitemOpen
  \bibinfo {note} {Molecules in an ensemble absorb independently if they are so dilute that there are no chemical, quantum statistical, or electromagnetic couplings. {For} electric dipole transitions, the imaginary part of the complex-valued dielectric permittivity, $\epsilon^{\prime \prime} (\nu, p, {T}) = {\Sigma} [^{eq}{N}_{R}(p,{T})\sigma _{{R} \to {S}}(\nu , p, {T})] c n(\nu ,p,{T})/(2\pi \nu)$, must be very much less than the absolute value of the real part: $\epsilon^{\prime \prime} (\nu ,p,{T}) << |\epsilon^{\prime} (\nu ,p,{T})|$. {Transition} cross sections proportional to $\nu ^{2}$ (or a higher even power) at low frequency guarantee that a low enough total absorber number density ${N}$ exists.}\BibitemShut {Stop}%
\bibitem [{RN3({\natexlab{i}})}]{RN3928}%
  \BibitemOpen
  \bibinfo {note} {Solute molecules must be much smaller than the wavelength of light for a solution to be uniform in the electromagnetic sense.}\BibitemShut {Stop}%
\bibitem [{RN3({\natexlab{j}})}]{RN32}%
  \BibitemOpen
  \bibinfo {note} {In {Fig.} \ref{fig:3}, the ${B}$ coefficient is determined from an oscillator strength $f_{{R} \to {S}} =$ 1 [see ref. \cite{RN1457,RN1458}] for the ${R}$ to ${S}$ transition. {Based} on the other parameters, the ${S}$ to ${R}$ transition has oscillator strength $f_{{S} \to {R}} =$ -1/2.}\BibitemShut {Stop}%
\bibitem [{RN3({\natexlab{l}})}]{RN3930}%
  \BibitemOpen
  \bibinfo {note} {In ref. \cite{RN3414}, the classical {Van Vleck-Weisskopf} formula may be found in {Eq.} (13-16), the quantum formula in {Eq.} (13-19), and comparison to experiment in {Fig.} 13-4. {Ref.} \cite{RN327} avoids critical $\nu_{ij}$ for $\nu $ substitution errors in ref. \cite{RN3500} to correctly obtain the $\nu^{\, 2}$ proportionality and to demonstrate that the quantum {Van Vleck-Weisskopf} lineshape unifies the low frequency {Debye} lineshape for dielectrics with the high frequency {Lorentzian} lineshape as limiting miscs. {The} quantum {Van Vleck-Weisskopf} lineshape is the same for absorption and stimulated emission, and the net absorption lineshape matches the classical average absorption in ref. \cite{RN3499}.}\BibitemShut {Stop}%
\bibitem [{RN3({\natexlab{m}})}]{RN3929}%
  \BibitemOpen
  \bibinfo {note} {{Van Vleck} called similar absorption minus stimulated emission terms "statistical" because they depend on two different statistical equilibrium initial band probabilities.}\BibitemShut {Stop}%
\bibitem [{RN3({\natexlab{n}})}]{RN3953}%
  \BibitemOpen
  \bibinfo {note} {If $g_{s} \neq g_{r}$, {Einstein's} special case of infinitely narrow lines gives $b_{r\to s} (\nu ,{T}) = b_{s\to r} (-\nu ,{T}) = 0$ at $h \nu = {\Delta} \mu _{r \to s}^{\text{o}}({T})$, thus making the forward and reverse ${B}$-coefficient spectra equal there as required by {Eq.} [\ref{eq21}].}\BibitemShut {Stop}%
\bibitem [{RN3({\natexlab{o}})}]{RN3948}%
  \BibitemOpen
  \bibinfo {note} {For intervalence-transfer absorption, the potential curves in {Fig.} \ref{fig:4} are "diabatic" potential curves that neglect the coupling between metal atoms. {Weak} coupling enables electron transfer between metal atoms near the diabatic curve crossing - see ref. \cite{RN3947}. {Such} electron transfer could prevent the equilibration needed for the near-vanishing reverse transition cross section in {Fig.} \ref{fig:4}b, but the coupling would hardly affect the equilibrium spectra for large solvent configuration displacement in {Fig.} \ref{fig:4}c.}\BibitemShut {Stop}%
\bibitem [{RN3({\natexlab{p}})}]{RN3955}%
  \BibitemOpen
  \bibinfo {note} {With static inhomogeneity, the {Stokes'} shift between ensemble absorption and ensemble emission can be less than that predicted by the single-molecule generalized {Einstein} relations -- see ref. \cite{RN2971}.}\BibitemShut {Stop}%
\end{thebibliography}

\end{document}